\documentclass[prl,reprint]{revtex4-1}
\usepackage{amsmath}
\usepackage{bbold}
\usepackage[utf8]{inputenc}
\usepackage[T1]{fontenc}
\usepackage[english]{babel}
\usepackage{verbatim}
\usepackage{textcomp}
\usepackage{ulem}
\usepackage[]{graphicx,xcolor}
\usepackage{soul}

\newcommand{\edit}[1]{{{\color{red}{#1}}}}

\newcommand{\p}{\partial}

\DeclareMathOperator{\sign}{sign}

\DeclareMathOperator{\diag}{diag}
\DeclareMathOperator{\rot}{rot}
\DeclareMathOperator{\pv}{\mathcal{P}}

\begin{document}
\title{
Dyakonov-like waveguide modes in an interfacial strip waveguide
}

\author{E.~V.~Anikin}
\email[]{e-mail: evgenii.anikin@skoltech.ru}
\affiliation{Skolkovo Institute of Science and Technology, 143025 Moscow Region, Russia}
\author{D.~A.~Chermoshentsev}
\affiliation{Skolkovo Institute of Science and Technology, 143025 Moscow Region, Russia}
\affiliation{Moscow Institute of Physics and Technology, 141701 Institutskiy pereulok 9, Moscow Region, Russia}
\affiliation{Russian Quantum Center, Skolkovo, Moscow 143025, Russia}
\author{S.~A.~Dyakov}
\email[]{e-mail: s.dyakov@skoltech.ru}
\affiliation{Skolkovo Institute of Science and Technology, 143025 Moscow Region, Russia}
\author{N.~A.~Gippius}
\affiliation{Skolkovo Institute of Science and Technology, 143025 Moscow Region, Russia}
\date{\today}

\begin{abstract}

We study Dyakonov surface waveguide modes in a waveguide represented by an interface of two anisotropic media confined between two air half-spaces. We analyze such modes in terms of perturbation theory in the approximation of weak anisotropy. We show that in contrast to conventional Dyakonov surface waves that decay monotonically with distance from the interface, Dyakonov waveguide modes can have local maxima of the field intensity away from the interface. We confirm our analytical results by comparing them with full-wave electromagnetic simulations. We believe that this work can bring new ideas in the research of Dyakonov surface waves.


\end{abstract}

\maketitle

Surface electromagnetic waves are solutions of Maxwell's equations in the form of monochromatic waves, which propagate along the interface of two dissimilar media and decay in the directions perpendicular to the interface. Examples of surface waves include surface plasmon polaritons at a metal-dielectric interface \cite{surface1982vm}, Tamm surface states at a photonic crystal boundary \cite{vinogradov06,dyakov2012surface,zhou2020theory}, surface solitons at a nonlinear interface \cite{kartashov2006surface} and many others. A special case of surface waves is Dyakonov surface waves (DSW) supported at the interface of two materials, at least one of which is anisotropic. Since the discovery in 1988 by M. Dyakonov  \cite{d1988new}, extensive research has been performed towards the theoretical study and experimental realization of DSWs and finding optimal material and geometrical configurations, which would be best suitable for potential practical implementations. Different combinations of isotropic, uniaxial, biaxial and chiral materials have been demonstrated to support Dyakonov surface waves at their interfaces \cite{averkiev1990electromagnetic, takayama2011dyakonov, gao2010dyakonov, agarwal2009theory, polo2007surface,zhou2020theory,repan2020wave, zhang2020unusual, fu2020complete,karpov2019dyakonov,li2020controllable,fedorin2019dyakonov, ardakani2016dyakonov, moradi2018terahertz,narimanov2018dyakonov}. The first experimental observation of Dyakonov surface waves using \textcolor{black}{the prism coupling method} was reported in 2009 by O. Takayama \cite{takayama2009observation}. Due to naturally small anisotropy of birefringent media, DSWs exist only in a narrow range of angular directions parallel to interface plane. Wider DSWs existence regions can be achieved using ultrathin partnering nanolayers, which can substantially release the Dyakonov condition and simplify DSWs experimental observation \cite{takayama2014lossless, sorni2015dyakonov}. Yet another realization of DSMs has been demonstrated theoretically and experimentally for metamaterials with artificially designed shape anisotropy \cite{artigas2005dyakonov, takayama2017midinfrared, jahani2016all, takayama2012practical, kajorndejnukul2019conformal, sorni2015dyakonov, takayama2017photonic,takayama2018experimental}. 

In contrast to surface plasmon-polaritons and many other surface waves, DSWs can exist at the interface of two  dielectrics. It means that DSWs potentially has no theoretical limit in propagation length. In this sense, a study of DSWs propagation in a corresponding waveguide would be interesting for the theory of nanophotonics and also rather intriguing from a practical viewpoint. Recently, waveguide properties of DSWs on cylindrical surfaces has been demonstrated in Ref.\,\cite{golenitskii2020dyakonov}. Due to the bending of the waveguide boundary, such DSMs have inevitable radiative losses, which tend to zero when the cylinder diameter tends to infinity. 

In this work, we consider a waveguide for DSWs, represented by a strip of the interface between two identical uniaxial birefringent dielectrics. We will demonstrate that such a flat waveguide can guide Dyakonov surface waves without losses.

\begin{figure}[b!]
    \centering
    \includegraphics[width=0.9\columnwidth]{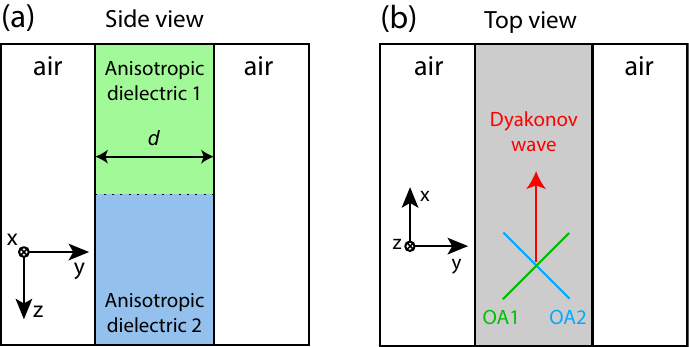}
    \caption{(a) Side view and (b) top view of Dyakonov waveguide. Optical axes (OA) of anisotropic dielectrics 1 and 2 are perpendicular to each other and form the angle of $45^\circ$ to the waveguide boundaries. }
    \label{fig:geometry}
\end{figure}

The schematic of the waveguide for DSW is shown in Fig.~\ref{fig:geometry}. It consists of two anisotropic dielectric slabs \textcolor{black}{separated from air by planes $|y| = d/2$. The slabs} have width $d$ in $y$-direction, are infinite in $x$-direction and semi--infinite in $z$ direction. These two dielectrics have different orientations of optical axes, and there is an interface between them at the plane $z = 0$. Although we are mostly interested in the case of two uniaxial media, we extend our scope to consideration of biaxial crystals with similar dielectric tensors whose axes are rotated by $45^\circ$ with respect to the initial coordinate system. Thus, the dielectric tensor inside the structure (at $|y| < d/2$) has form
\begin{equation}
    \begin{gathered}
    \hat{\epsilon}(z) = \hat{\epsilon}_0 + \delta\hat{\epsilon}(z), \,\,\,\,\,
    \hat{\epsilon}_0 = \diag(\epsilon_1, \epsilon_1, \epsilon_2)\\
    \delta\hat{\epsilon}(z) = \sign(z)
                    \begin{pmatrix}
                        0 & \delta\epsilon & 0\\
                        \delta\epsilon & 0 & 0\\
                        0 & 0 & 0
                    \end{pmatrix}
    \end{gathered}
\end{equation}
We treat the parameters $\epsilon_{1,2}$ and $\delta\epsilon$ as independent ones. However, by setting $\epsilon_1 - \epsilon_2 = \delta\epsilon$ one can consider the case of two uniaxial crystals. \textcolor{black}{Please note that such a representation of dielectric tensor $\hat{\epsilon}$ is possible only for the optical axes rotation angles of $\pm45^\circ$ relative to the $x$-axis.}

In such a structure, there can exist lossless modes propagating along the $x$ axis, which are localized in $z$ direction near the interface of two slabs. We will call them Dyakonov waveguide modes (DWM). 
DWMs can be analyzed in the framework of perturbation theory by the nondiagonal part of the permittivity tensor. Without perturbation, the waveguide is translationally invariant in both $x$ and $z$ directions and all eigenmodes have form of plane waves $\vec{E}_{k_x,k_z}(y)e^{ik_x x +ik_zz}$, where $k_x$ and $k_z$ are the projections of the wavevector $\vec{k}$. When the perturbation is present, a DWM can appear with a lower frequency than all waveguide modes at a fixed $k_x$. For weak perturbation, DWM decays slowly away from the interface. Because of this, it is possible to describe DWMs in terms of unperturbed waveguide modes multiplied by a slowly varying envelope.

\begin{figure*}
    \includegraphics[width=1\linewidth]{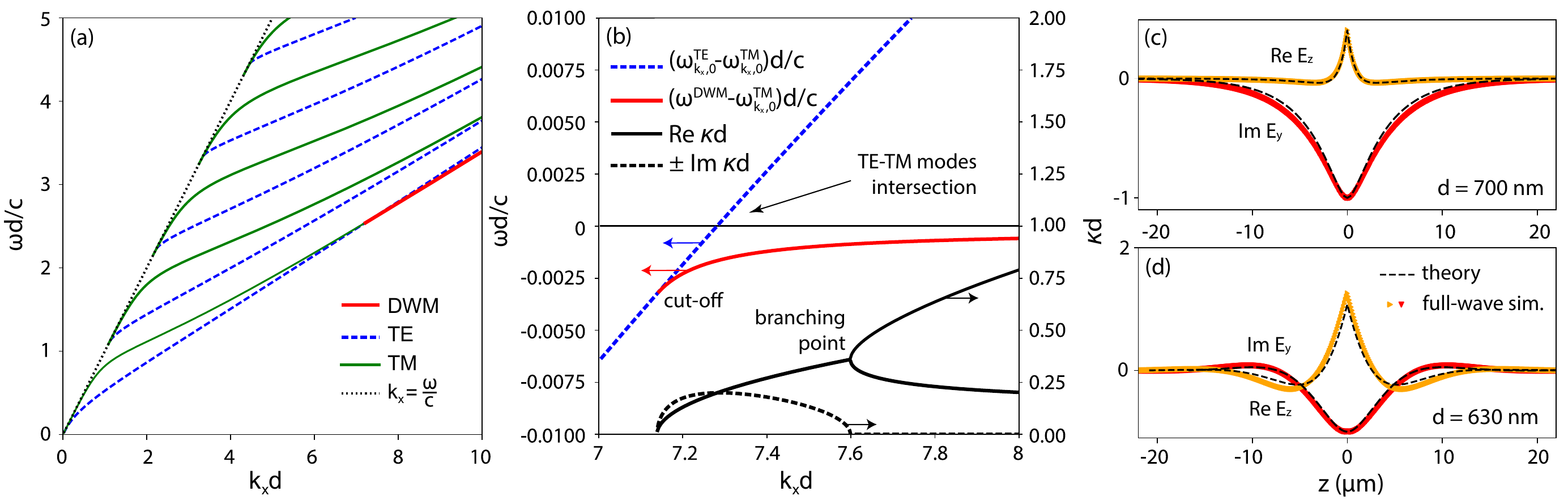}
    \caption{\textcolor{black}{(a) The dispersions of the DWM (red solid line) and the TE and TM modes of an anisotropic waveguide (blue dashed line and green solid line) with permittivity $\hat{\epsilon}_0$ travelling along $x$ axis. (b) The in-plane wavevector dependence of the real and imaginary parts of the decay constants $\kappa_{1,2}$ (black solid and dashed line). The dispersions of the DWM (red line) and the TE waveguide mode of an anisotropic waveguide (blue dashed line) with permittivity $\hat{\epsilon}_0$ shown as a difference with the dispersion of the TM waveguide mode. In the presented diagram, the TM waveguide mode corresponds to the horizontal line $\omega d/c = 0$ (black thin solid line). (c), (d) Theoretically calculated fields $E_y(0,z)$ and $E_z(0,z)$ (black dashed lines) shown together with COMSOL simulation results (red and yellow lines) for $\lambda=1550$\,nm and (c) $k_xd$ = 7.262, $d=700$\,nm, (d) $k_xd$ = 8.1938, $d=630$\,nm. All calculations are made for $\delta\epsilon=0.5$,  $\epsilon_1 = 9.5$, $\epsilon_2 = 9$.}}
    \label{fig:234}
\end{figure*}

The general solution for waveguide modes in an anisotropic waveguide with permittivity $\hat{\epsilon}_0$ can be obtained analytically for arbitrary $k_x$ and $k_z$ (see Supplemental Materials at~\footnote{See Supplemental materials at http://link.aps.org/supplemental/xxx for the analytical solution for waveguide modes in an anisotropic waveguide.}). For waves propagating in $x$ direction, the modes are identical to those of an isotropic waveguide, i.e. they have the same fields and frequency. Namely, the modes with $\vec{E} \parallel Oz$ coincide with TE modes of an isotropic waveguide with permittivity $\epsilon_2$, and the modes with $\vec{E} \perp Oz$ coincide with TM modes of an isotropic waveguide with permittivity $\epsilon_1$. For $\epsilon_2 < \epsilon_1$, two lowest waveguide modes intersect (have the same frequency  $\omega$, see Fig.~\ref{fig:234}a) at some value of propagation constant $k_x$. Thus, both of them should be taken into account in the decomposition of DWMs over waveguide eigenmodes. 

Closely to the intersection of fundamental TE and TM modes, the nondiagonal perturbation $\delta\hat\epsilon$ leads to considerable mixing of these modes. Moreover, the contribution of all other modes can be neglected provided that the mode spacing (which has order $c/d$) is much larger than the distance between the two lowest modes.

We consider Maxwell's equations as an eigenvalue problem
\begin{equation}
    \begin{gathered}
    \label{eig_maxwell_eq}
    c^2\rot\rot{\vec{E}} = \omega^2 \hat\epsilon\vec{E}
    \end{gathered}
\end{equation}
where $\omega$ is the frequency of electromagnetic oscillations \textcolor{black}{and $c$ is the speed of light}. All eigenmodes of this problem constitute a basis in the space of fields. 
As the unperturbed problem has translational symmetry in $z$ direction, we will consider the basis
\begin{equation}
    \vec{\mathcal{E}}^{n}_{k_x,k_z}(y,z) = \vec{E}^{n}_{k_x,k_z}(y)e^{ik_{z}z},
\end{equation}
with corresponding eigenvalues $\omega_{k_x,k_z,n}$, where $n$ is the index enumerating different modes. The set of all modes at particular $k_x$, $k_z$ includes a finite number of waveguide modes which decay exponentially away from the waveguide and a continuum of free space modes corresponding to plane wave scattering on the waveguide. Below, we don't take the continuum modes into account because, as it will be shown, only the modes with lowest frequencies contribute to DWM, and the expansion including lowest TE and TM modes reads
\begin{equation}
    \label{field_expansion}
    \vec{E}(y,z) = \int \frac{dk_z}{2\pi} \left[ \alpha(k_z) \vec{\mathcal{E}}^{\mathrm{\scriptscriptstyle TE}}_{k_x,k_z} + 
                \beta(k_z) \vec{\mathcal{E}}^{\mathrm{\scriptscriptstyle TM}}_{k_x,k_z} \right]
\end{equation}
Taking into account the orthogonality of eigenvectors of the
eigenvalue problem \eqref{eig_maxwell_eq}, we normalize the waveguide modes by a condition
\begin{equation}
   \int (\vec{\mathcal{E}}^{n}_{k_x,k_z}, \hat{\epsilon}_0 \vec{\mathcal{E}}^{n'}_{k_x,k'_z}) \,dydz = 
    2\pi\delta(k_z-k'_z)\delta_{nn'}.
\end{equation}

Below we obtain the equation for the envelopes $\alpha(k_z)$ and $\beta(k_z)$. Due to small $\delta \epsilon$, DWM should have slow field dependence on $z$, and, hence, the field envelopes $\alpha(k_z)$ and $\beta(k_z)$ are significant at small $k_z$ only. After substituting this expansion in Maxwell equations and taking scalar product of both sides of these equations by $\vec{\mathcal{E}}^{\mathrm{\scriptscriptstyle TE}}_{k_x,k_z}$ and $\vec{\mathcal{E}}^{\mathrm{\scriptscriptstyle TM}}_{k_x,k_z}$, one obtains
\begin{equation}
    \label{eff_eq_0}
    \begin{gathered}
    \gamma^{\mathrm{\scriptscriptstyle TE}}_{k_x,k_z}\alpha(k_z) = 
        \omega^2\int\frac{dk'_z}{(2\pi)} 
        \beta(k_z') 
        \langle\vec{\mathcal{E}}^{\mathrm{\scriptscriptstyle TE}}_{k_x,k_z}\delta\hat\epsilon \vec{\mathcal{E}}^{\mathrm{\scriptscriptstyle TM}}_{k_x,k_z'}\rangle\\
   \gamma^{\mathrm{\scriptscriptstyle TM}}_{k_x,k_z}  \beta(k_z) = 
        \omega^2\int\frac{dk'_z}{(2\pi)} \alpha(k_z') 
        \langle\vec{\mathcal{E}}^{\mathrm{\scriptscriptstyle TM}}_{k_x,k_z}\delta\hat\epsilon \vec{\mathcal{E}}^{\mathrm{\scriptscriptstyle TE}}_{k_x,k_z'}\rangle
    \end{gathered}
\end{equation}
where $\gamma_{k_x,k_z}^{\mathrm{\scriptscriptstyle TE(TM)}}=\left(\omega_{k_xk_z}^{\mathrm{\scriptscriptstyle TE(TM)}}\right)^2 - \omega^2$.

The matrix elements
$\langle\vec{\mathcal{E}}^{\mathrm{\scriptscriptstyle TE}}_{k_x,k_z}\delta\hat\epsilon \vec{\mathcal{E}}^{\mathrm{\scriptscriptstyle TE}}_{k_x,k_z'}\rangle$ and
$\langle\vec{\mathcal{E}}^{\mathrm{\scriptscriptstyle TM}}_{k_x,k_z}\delta\hat\epsilon \vec{\mathcal{E}}^{\mathrm{\scriptscriptstyle TM}}_{k_x,k_z'}\rangle$
are not present in Eq.~\eqref{eff_eq_0} because they
turn out to be zero due to symmetry properties of TE and TM modes.
Also, at small wavevectors 
the matrix element corresponding to mixing between TE and TM modes
$\langle\vec{\mathcal{E}}^{\mathrm{\scriptscriptstyle TE}}_{k_x,k_z}\delta\hat\epsilon \vec{\mathcal{E}}^{\mathrm{\scriptscriptstyle TM}}_{k_x,k_z'}\rangle$ reads
\begin{multline}
    \label{matrix_element_kz}
\langle\vec{\mathcal{E}}^{\mathrm{\scriptscriptstyle TE}}_{k_x,k_z}\delta\hat\epsilon \vec{\mathcal{E}}^{\mathrm{\scriptscriptstyle TM}}_{k_x,k_z'}\rangle 
    =2i\pv{\left(\frac{1}{k_z-k_z'}\right)}k_z\delta\epsilon\sigma,\\
    \sigma\delta\epsilon =
    \left.
    \int_{-a/2}^{a/2} dy
            \left(\p_{k_z}\vec{E}^{\mathrm{\scriptscriptstyle TE}}_{k_x,k_z}(y)\delta\hat{\epsilon}\vec{E}^{\mathrm{\scriptscriptstyle TM}}_{k_x,k'_z}(y)\right) 
    \right|_{k_z,k'_z=0}    
\end{multline}
where $\pv$ denotes the principal value.
After substituting \eqref{matrix_element_kz} to \eqref{eff_eq_0} and performing the inverse Fourier transform of 
\eqref{eff_eq_0}, one gets
a system of ODEs in coordinate space for Fourier images of $\alpha(k_z)$ and $\beta(k_z)$, $\alpha(z)$ and $\beta(z)$.

Expanding the frequencies of TE and TM modes in $k_z$ up to quadratic terms, one gets
\begin{equation}
    \label{eff_eq_on_envelopes}
    \begin{pmatrix}
        \gamma^{\mathrm{\scriptscriptstyle TE}}_{k_x,0}- \frac{1}{2m_{1}}\frac{\p^2}{\p z^2} & 
        i\omega^2\sigma\delta\epsilon\p_z \sign(z)\\
        i\omega^2\sigma\delta\epsilon\sign(z)\p_z & 
        \gamma^{\mathrm{\scriptscriptstyle TM}}_{k_x,0} - \frac{1}{2m_{2}}\frac{\p^2}{\p z^2}
    \end{pmatrix}
    \begin{pmatrix}
        \alpha(z)\\
        \beta(z)
    \end{pmatrix}=0
\end{equation}
where all the coefficients in \eqref{eff_eq_on_envelopes} are expressed through the quantities referring to the planar waveguide:
\begin{equation}
    \label{eff_coefs}
    \begin{gathered}
        m_{1(2)}^{-1} = \left.\frac{\p^2}{\p k_z^2}\left(\omega_{k_x,k_z}^{\mathrm{\scriptscriptstyle TE(TM)}}\right)^{2}\right|_{k_z=0}
    \end{gathered}
\end{equation}
and $\sigma$ is defined by Eq.~\eqref{matrix_element_kz}.
\textcolor{black}{The system \eqref{eff_eq_on_envelopes}  have solutions in the form of decaying exponential functions for $z>0$ and $z<0$ which can be matched using the boundary conditions at $z=0$. These exponentially decaying solutions exactly correspond to DWMs localized near the interface with field distributions given by Eq.~\eqref{field_expansion}.} By integrating \eqref{eff_eq_on_envelopes} in the neighborhood of $z = 0$, one gets that $\p_z \beta(z)$ is continuous, 
and the condition for $\p_z\alpha(z)$ reads
\begin{equation}
    \p_z\alpha(+0) - \p_z\alpha(-0) = 
    4im_1\omega^2\sigma\delta\epsilon
    \beta(0).
\end{equation}

The explicit expressions for coefficients \eqref{eff_coefs} in the effective equation for envelopes \eqref{eff_eq_on_envelopes} can be easily calculated numerically. They also drastically simplify in the limit when $\epsilon_1 - \epsilon_2 \ll \epsilon_1$ and all modes are very close to that of the planar isotropic waveguide. In this limit, fundamental TE and TM modes intersect at $k_x d \gg 1$. So, one can utilize the large--wavevector expansion of $\omega^{\mathrm{\scriptscriptstyle TE}}_{k_x,0}$ and $\omega^{\mathrm{\scriptscriptstyle TM}}_{k_x,0}$ to find the intersection point. The resulting wavevector of intersection reads
\begin{equation}
    k_x d \approx \sqrt{\epsilon_1}
    \left(\frac{2\pi^2\sqrt{\epsilon_1-1}}
        {\epsilon_1(\epsilon_1-\epsilon_2)}\right)^\frac{1}{3}.
\end{equation}
At large $k_x$, the parameters $m_1^{-1}$ and $m_2^{-1}$ approach $c^2/\epsilon_1$, and the asymptotic behavior of the matrix element is $\sigma = (k_x\epsilon_1)^{-1}$.

The examination of the system \eqref{eff_eq_on_envelopes} allows finding the domain of existence, the dispersion law, and the field structure of DWMs. Before we go in further detail, let us emphasize that such modes propagate along $x$ axis completely without losses. \textcolor{black}{This is because, } as shown below, the frequencies of DWMs in such structure are lower than the frequencies of the waveguide modes. Thus, DWM cannot scatter into waveguide modes or free space modes without violation of energy or momentum conservation law. Therefore, the imaginary parts of the frequency and the wavevector are exactly zero, $\omega'' = 0$ and $k_x'' = 0$.
\begin{figure*}[t!]
    \centering
    \includegraphics[width=0.8\textwidth]{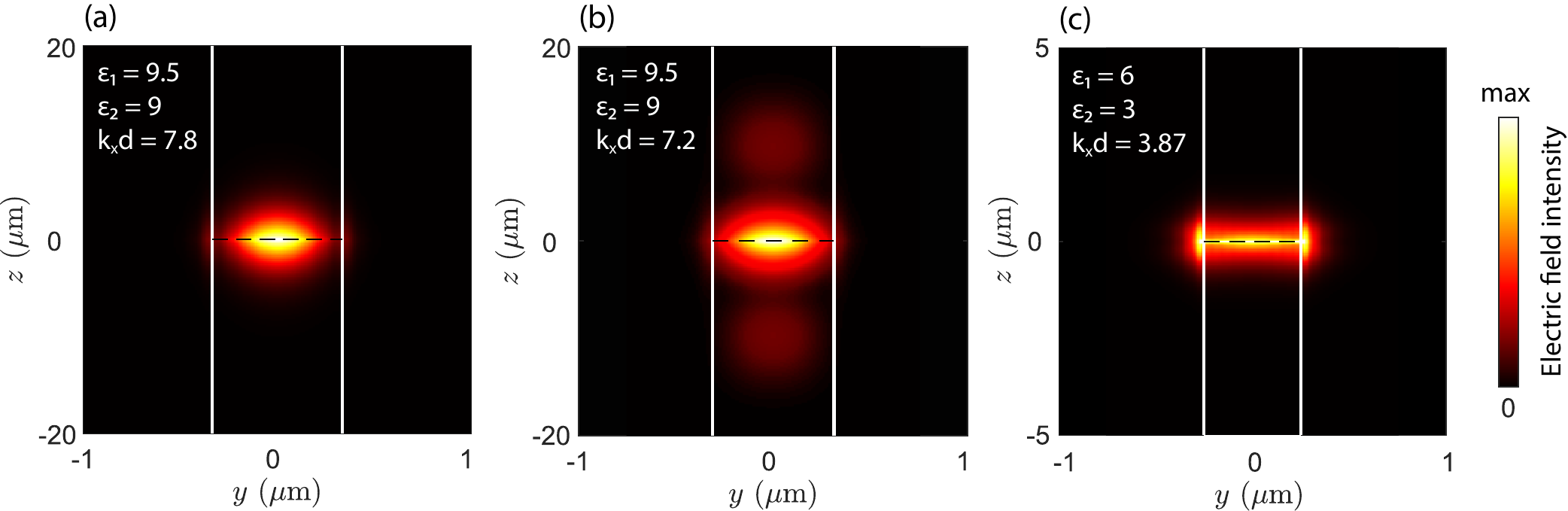}
    \caption{Electric field intensity in DWMs at $\lambda=1.55$\,$\mu$m and (a) $\varepsilon_1=9.5$, $\varepsilon_2=9$, $k_xd=7.8$, $d\approx670$\,nm, (b) $\varepsilon_1=9.5$, $\varepsilon_2=9$, $k_xd=7.2$, $d\approx625$\,nm  (c) $\varepsilon_1=6$, $\varepsilon_2=3$, $k_xd\approx3.87$, $d=500$\,nm. White vertical lines denote the waveguide boundaries, black dashed horizontal lines denote the interface between anisotropic dielectrics.}
    \label{fig:int}
\end{figure*}

The exponentially decaying solution of \eqref{eff_eq_on_envelopes} obeying the boundary conditions reads
\begin{equation}
    \label{exact_envelopes}
    \begin{gathered}
        \alpha(z) =\frac{\omega^2\sigma\delta\epsilon e^{-\kappa_2|z|}}
            {\left( \gamma^{\mathrm{\scriptscriptstyle TE}}_{k_{x},0} - \frac{\kappa_2^2}{2m_{1}}\right) }
                     -   \frac{\omega^2\sigma\delta\epsilon e^{-\kappa_1|z|}}
            {\left(\gamma^{\mathrm{\scriptscriptstyle TE}}_{k_{x},0} - \frac{\kappa_1^2}{2m_{1}}\right)}\\             
        \beta(z) = -i\left(\frac{e^{-\kappa_2|z|}}{\kappa_2} - 
            \frac{e^{-\kappa_1|z|}}{\kappa_1})
            \right)
    \end{gathered}
\end{equation}

where $\kappa_{1,2}$ are two roots with positive real
part of the characteristic equation
\begin{equation}
    \label{char_eq}
    \left(\gamma^{\mathrm{\scriptscriptstyle TE}}_{k_{x},0} - \frac{\kappa^2}{2m_{1}}\right) \left(\gamma^{\mathrm{\scriptscriptstyle TM}}_{k_{x},0} - \frac{\kappa^2}{2m_{2}}\right) + \omega^4\sigma^2\delta\epsilon^2\kappa^2 = 0
\end{equation}
The implicit dispersion follows from boundary conditions and has the form
\begin{multline}
    \label{dm_dispersion}
    \sqrt{m_1m_2\gamma^{\mathrm{\scriptscriptstyle TE}}_{k_{x},0}\gamma^{\mathrm{\scriptscriptstyle TM}}_{k_{x},0}}
    = 2m_1m_2(\omega^2\sigma\delta\epsilon)^2 -m_2\gamma^{\mathrm{\scriptscriptstyle TM}}_{k_{x},0}
\end{multline}
Whether there exists a solution for these equations, depends on the relation between $\omega^{\mathrm{\scriptscriptstyle TE}}_{k_x,0}$ and $\omega^{\mathrm{\scriptscriptstyle TM}}_{k_x,0}$ and the off--diagonal matrix element. In particular, the solution of Eq.~\eqref{dm_dispersion} exists when the lower cutoff condition is satisfied \textcolor{black}{(see red curve in Fig.\,\ref{fig:234}b)}:
\begin{equation}
(\omega^{\mathrm{\scriptscriptstyle TM}}_{k_x,0})^2 - (\omega^{\mathrm{\scriptscriptstyle TE}}_{k_x,0})^2 \le 2m_1(\omega^{\mathrm{\scriptscriptstyle TE}}_{k_x,0})^4\sigma^2\delta\epsilon^2
\end{equation}

From expressions \eqref{exact_envelopes}, \eqref{char_eq}, \eqref{dm_dispersion} one can see that the decay constants $\kappa_{1,2}$ can be either complex and conjugate to each other or both real. The case of complex $\kappa_{1,2}$ implements near the vicinity of the TE and TM modes intersection, \textcolor{black}{whereas f}ar enough from the intersection point, both $\kappa_{1}$ and $\kappa_{2}$ are real. These alternatives are separated by the branching point where two roots of the characteristic equation \eqref{char_eq} coincide. \textcolor{black}{As the DWMs} are qualitatively different for these two cases \textcolor{black}{let us consider them in more detail.}
 
\textcolor{black}{Firstly, we consider a large separation between TE and TM modes. In this case, f}or the solutions of the characteristic equation and the field distributions, simple asymptotic expressions can be obtained. \textcolor{black}{Thereby,} the DWM frequency \textcolor{black}{ evaluated} from \eqref{dm_dispersion} is very close to TM mode frequency:
\begin{equation}
    \label{approx_domega}
    \begin{gathered}
    \delta\omega = \omega^{\mathrm{\scriptscriptstyle TM}}_{k_x,0} - \omega^{\mathrm{\scriptscriptstyle DWM}} \approx \frac{2m_1m_2(\omega^{\mathrm{\scriptscriptstyle TM}}_{k_x,0})^7\sigma^4\delta\epsilon^4}
    {(\omega^{\mathrm{\scriptscriptstyle TE}}_{k_x,0})^2 - (\omega^{\mathrm{\scriptscriptstyle TM}}_{k_x,0})^2} \propto \delta\epsilon^4
    \end{gathered}
\end{equation}

Also, the amplitudes of TE and TM modes in this limit take the form
\begin{equation}
    \label{approx_solution}
    \begin{gathered}
    \alpha(z) = 
       \frac{2m_1(\omega^{\mathrm{\scriptscriptstyle TM}}_{k_x,0})^2
              \sigma\delta\epsilon}{\kappa_1}
                    e^{-\kappa_1|z|},\,\,\,~ 
                    \beta(z) = -ie^{-\kappa_2|z|}
    \end{gathered}
\end{equation}
where the decay constants read $\kappa_1^2 \approx 2m_1\left[(\omega^{\mathrm{\scriptscriptstyle TE}}_{k_x,0})^2 - (\omega^{\mathrm{\scriptscriptstyle TM}}_{k_x,0})^2\right]$ and $\kappa_2^2 \approx 4m_2 \omega^{\mathrm{\scriptscriptstyle TM}}_{k_x,0}\delta\omega$. Thus, in the considered limits the main contribution to DWM is from the TM waveguide mode. Its amplitude is slowly decaying with the decay length $\sim 1/\kappa_2$ \textcolor{black}{ (Fig.\,\ref{fig:int}a)}, whereas the amplitude of the TE mode is small by $\delta\epsilon$, and the TE mode is localized near the interface on a much shorter length $\sim 1/\kappa_1$. The theoretically calculated profiles of $E_y$ and $E_z$ for the considered case demonstrate excellent agreement of with the results of full-wave electromagnetic simulations of DWMs made in COMSOL (Fig.\,\ref{fig:234}c).

 Now let us analyze the vicinity of TE and TM modes intersection when the approximation of Eqs.~\eqref{approx_domega} and \eqref{approx_solution} becomes not valid. The exact solution of Eqs.\,\eqref{exact_envelopes}, \eqref{char_eq}, \eqref{dm_dispersion} should be utilized in this case, and the contributions of TE and TM modes to DWM as well as the inverse decay lengths $\kappa_1$ and $\kappa_2$, become comparable. Also, the difference between DWM frequency and the lowest TM mode frequency  $\delta\omega$ becomes proportional to $\delta\epsilon^2$, so the separation between DWM and waveguide modes is maximal near the TE and TM modes intersection. As it was stated before, the inverse decay lengths $\kappa_1$ and $\kappa_2$ acquire nonzero imaginary part closely to the modes intersection (see Fig.~\ref{fig:234}b), so the fields exhibit oscillations with $z$ which results in additional local maxima of the field intensity at some distance from the interface (see Fig.\,\ref{fig:234}d and Fig.\,\ref{fig:int}b). This distinct feature distinguishes DWMs from conventional Dyakonov surface waves which exist at the infinite flat interface and decay monotonically.

It should be emphasized once again that the two-mode approximation is valid only provided that the difference between TE and TM modes frequencies is much less then the mode spacing, $\omega^{\mathrm{\scriptscriptstyle TE}}_{k_x,0} - \omega^{\mathrm{\scriptscriptstyle TM}}_{k_x,0} \ll \frac{\pi^2}{k_xd^2\epsilon_{1,2}}$. As $\omega^{\mathrm{\scriptscriptstyle TE}}_{k_x,0} - \omega^{\mathrm{\scriptscriptstyle TM}}_{k_x,0}$ grows with $k_x$, the two--mode approximation cannot be applied far from the mode intersection point. The above analytical considerations only apply to the case of a small anisotropy $\delta\varepsilon\ll\varepsilon_1,\varepsilon_2$. Properties of DWMs in a more general case of an arbitrary anisotropy, their domain of existence, the presence of a branching point are subjects of separate research. However, in Fig.\,\ref{fig:int}c we show an example of the DWM obtained numerically in COMSOL for $\varepsilon_1=6$ and $\varepsilon_2=3$. One can see that the electric field intensity decays with distance from the interface even faster than in the considered cases of small anisotropy.

\textcolor{black}{Finally, the considered DWMs can be generalized to a non-45$^\circ$ rotation of optical axes as well as to other types of partnering media including different combinations of isotropic, uniaxial, biaxial, chiral materials and also photonic crystals. Existence of DWMs in each particular case is a subject of separate research.}

In conclusion, we have analytically and numerically demonstrated the existence of Dyakonov waveguide modes which can propagate without losses at the interface of two anisotropic dielectric waveguides. On the dispersion diagram, DWMs exist near the intersection of the lowest TE and TM modes of \textcolor{black}{these} waveguides. We have shown that DWMs generally are localized on the interface but under certain conditions, they also may have additional local maxima of the field intensity at some distance from the interface.

\begin{acknowledgments}
This work was supported by the Russian Foundation for Basic Research (Grant \textnumero 18-29-20032).
\end{acknowledgments}

\end{document}